\documentclass[preview,5p,times,twocolumn]{elsarticle}

\usepackage{graphicx}

\usepackage{amssymb}
\usepackage{amsthm}

\usepackage{amsmath}
\usepackage{siunitx}
\usepackage{hyperref}
\usepackage{cleveref}
\usepackage{algorithm}
\usepackage{algpseudocode}
\usepackage[normalem]{ulem}
\usepackage{dblfloatfix}
\usepackage[dvipsnames]{xcolor}

\newcounter{bla}

\journal{Computer Physics Communications}

\begin{document}

\begin{frontmatter}



\title{\texorpdfstring{$\pi$-PIC}{pi-PIC}: a modular framework for particle-in-cell developments and simulations}


\author[a]{Frida Brogren\corref{author}}
\author[a,b]{Christoffer Olofsson}
\author[a]{Joel Magnusson}
\author[a,c]{Mattias Marklund}
\author[a]{Arkady Gonoskov}

\cortext[author] {Corresponding author.\\\textit{E-mail address:} frida.brogren@physics.gu.se}
\address[a]{Department of Physics, University of Gothenburg, SE-412 96 Gothenburg, Sweden}
\address[b]{Department of Information Technology, University of Borås, SE-501 90 Borås, Sweden}
\address[c]{Chalmers University of Technology, Department of Physics, SE-412 96 Gothenburg, Sweden}

\begin{abstract}

Particle-in-cell (PIC) codes are indispensable tools for studying plasma-based interactions across a wide range of physical regimes. The predictive capabilities of these codes have dramatically increased by continuous advances in algorithms and computing hardware. Nevertheless, greater computational performance and broader framework capabilities often lead to increased software complexity, hindering the adoption, evaluation, and comparison of new numerical developments. In this work, we present a Python-controlled framework designed to facilitate the dissemination and adoption of novel PIC developments by providing a unified interface for accommodation, cross-testing, and comparison of PIC algorithms. To demonstrate the flexibility of the proposed interface, we implement and discuss a range of PIC schemes together with applications involving absorbing boundary layers, moving-window simulations, and tight laser focusing. 



   \end{abstract}
\end{frontmatter}

\section{Introduction}

In recent years, impressive progress has been made in developing advanced schemes that overcome or mitigate fundamental limitations of the particle-in-cell (PIC) method. Significant effort has been dedicated to creating faster, more versatile, exact and user-friendly PIC codes. Some of the most widely used codes are PIConGPU \cite{burau_picongpu_2010,burau_picongpu_2010_code}, Smilei \cite{derouillat_smilei_2018, derouillat_smilei_2018_code}, Epoch \cite{Epoch}, Osiris \cite{fonseca_osiris_2002, fonseca_osiris_2002_code}, PICADOR \cite{bastrakov_particle_cell_2012} and the codes included in the BLAST project (e.g., WarpX \cite{vay_warp-x_2018, vay_warp-x_2018_code}, WakeT \cite{Ferran_Pousa_2019}, HiPACE \cite{DIEDERICHS2022108421} and FBPIC \cite{lehe_spectral_2016, lehe_spectral_2016_code}). Codes like Smilei, Osiris, Epoch and WarpX target a wide user base with a large variety of functionality suitable for different types of interactions, for example ionization, Breit-Wheeler pair creation, Schwinger processes, radiation and ionization reactions,  as well as collisions. Other codes, including FBPIC, WakeT, and HiPACE, aim to enhance speed through different strategies, such as employing simplified geometries, adopting quasi-static methods, and utilizing boosted frame simulations.

Nevertheless, as the PIC community grows and the applicability of PIC codes widens, the need for testing and incorporating new physics and algorithms persists. Additionally, as these codes evolve, several with a focus on optimized execution, modifying them becomes more complex and expensive from a developer's perspective. In other areas of computational physics and applied mathematics, such challenges have been alleviated by standard interfaces that enable testing and comparison of different methods within a single design pattern (e.g., OpenFOAM \cite{OpenFOAM}, ASE \cite{ASE}, LAMMPS \cite{LAMMPS,LAMMPS_code}, BLAS \cite{BLAS}, LAPACK \cite{LAPACK,LAPACK_code}, FFTW \cite{FFTW,FFTW_code}, COIN-OR \cite{COIN_OR} and Scikit-learn \cite{sklearn,sklearn_code}).

Certainly, the developers of almost every well-established code strive to cover such functionality, but for the wider PIC community, such an interface remains to be established. A code suitable for cross-community collaboration should satisfy three key requirements: low knowledge barrier for contributions, high flexibility to accommodate the wide range of projects within the PIC community, and decentralized development processes. In this article, we present a code framework, $\pi$-PIC, which attains these qualities by featuring a modular structure, that enables the independent development of PIC schemes and extensions. This facilitates the straightforward integration of new physical processes into existing PIC algorithms. To further simplify usage, the simulations are initialized and advanced through a standard Python interface. To not compromise computational time, extensions and PIC schemes are pre-compiled and multi-threaded. The capabilities of the devised interface are shown by considering a number of PIC implementations, extensions and Python scripts that combine them for validation tests.

The article is arranged as follows: in \cref{sec:lim_and_prospects}, we give a brief overview of selected PIC algorithms, numerical inaccuracies connected to the methods and future prospects. \Cref{sec:arch} considers architecture and utilization of the presented framework. The subsequent sections illustrate the use of extensions and solvers. Extensions for handling tight focusing of laser pulses and absorbing boundaries are presented in \cref{sec:appl}, while \cref{sec:solvers} focuses on the development and benchmarking of numerical solvers. Specifically, \cref{sec:two_stream_benchmark} presents a benchmark of explicit and implicit electrostatic solvers using the two-stream instability, while \cref{sec:benchmark} compares an energy-conserving solver implemented in $\pi$-PIC with the Smilei PIC code \cite{derouillat_smilei_2018_code} in the context of laser–wakefield acceleration.

\section{Limitations and prospects for the PIC method}\label{sec:lim_and_prospects}

The challenges of simulating plasmas typically revolve around reducing numerical artifacts and improving computational efficiency. Over the years, a variety of PIC formulations have been developed to address these challenges.

PIC methods may be distinguished according to whether they are local or global, explicit or implicit, and according to whether fields are represented in coordinate or spectral space. While these distinctions remain useful, many modern PIC formulations are additionally characterized by their particle--field coupling strategy and field representation. In particular, PIC methods differ in how particles and fields are coupled, how electromagnetic fields are represented, and how the computational mesh is discretized. Examples range from conventional particle push–deposition–field update schemes to coupled particle–field formulations, from electromagnetic field-based to potential-based descriptions, and from structured Cartesian grids to staggered, finite-element, and unstructured mesh representations.

Owing to the high computational cost of plasma simulations, the development of PIC methods has historically been driven by computational efficiency, often favoring inexpensive algorithms over more accurate but computationally demanding alternatives. Due to their parallelization capability and fast execution, explicit local methods typically achieve higher computational speeds than other approaches. Moreover, the rapid growth of parallel computing resources has driven a broad preference for explicit methods. 

The choice of PIC formulation influences not only computational efficiency, but also which physical and mathematical properties of the continuous system are preserved by the discretization. In the following we discuss three examples where numerical artifacts may arise from the failure to preserve such properties: (1) numerical dispersion, (2) charge conservation, and (3) energy and momentum conservation.

(1) Numerical dispersion emerges as a result of the discretization of electromagnetic waves, causing the discretized waves to travel at velocities that differ from the speed of light. For example, the popular Maxwell solver Finite-Difference Time-Domain (FDTD) \cite{yee_numerical_1966}, which is a local explicit method, exhibits direction-dependent numerical dispersion. Several modifications of the FDTD scheme have been proposed to reduce numerical dispersion, eliminating it completely in certain directions while mitigating it in others \cite{pukhov_three-dimensional_1999,karkkainen_low_dispersion_2006,cole_high-accuracy_1997,lehe_numerical_2013,cowan_generalized_2013}. Alternatively, spectral solvers can eliminate numerical dispersion entirely, although at the cost of introducing global operations \cite{birdsall_plasma_2018,vay_domain_2013}. The additional computational cost can be alleviated through parallel implementations based on domain decomposition \cite{vay_domain_2013}.

(2) Apart from low numerical dispersion, another desirable property of PIC codes is charge conservation. In the advancement of the fields, charge conservation is fulfilled if Ampère-Maxwell's law and Gauss's law are simultaneously satisfied. Ampère-Maxwell's law is commonly used to advance the field in time, whereas Gauss's law is not involved directly in the evolution of the state. Consequently, one common tactic to ensure charge conservation in explicit PIC is to correct the fields retroactively using Gauss's law \cite{marder_method_1987,munz_divergence_2000}. Another option is to employ current deposition schemes that ensure local charge conservation \cite{villasenor_rigorous_1992,esirkepov_exact_2001, umeda_new_2003, kong_particle--cell_2011,steiniger_ez_2023}. It has also been shown that similar local charge conserving deposition schemes can be implemented on general finite-element meshes \cite{pinto_charge-conserving_2014}.

Local charge conserving deposition schemes can be interpreted as enforcing discrete analogues of the conservation laws satisfied by the continuous system. More recently, this notion has been directly realized in variational PIC methods where the particle and field evolution is derived from a discrete action which respects a discrete gauge invariance, leading to exact preservation of Gauss's law \cite{squire_geometric_2012}. Similarly, potential-based formulations employing generalized momenta have demonstrated exact charge conservation \cite{christlieb_particle--cell_i_2025,christlieb_particle--cell_ii_2024,christlieb_particle--cell_iii_2025}. 

(3) Furthermore, standard explicit PIC implementations do not generally conserve energy or momentum exactly. This may result in numerical heating, particularly when the Debye length is under-resolved, as aliased modes can appear as artificial thermal fluctuations \cite{pukhov_particle_in_cell_2016, powis_accuracy_2024, langdon_effects_1970}. One way of mitigating these effects is through smoother particle shape functions, which act as low-pass filters for short-wavelength oscillations \cite{birdsall_plasma_2018}. Another approach is to construct the particle and field updates such that a discrete analogue of the total energy is preserved exactly, such as the explicit method implemented in $\pi$-PIC \cite{gonoskov_jcp_2024} or the recently proposed explicit energy-conserving scheme of Ricketson and Hu \cite{ricketson.jpc.2025}. 

Similarly as for charge conservation there also exists a growing class of PIC methods which conserves energy through deriving particle updates, field updates, and conservation properties from a discretization that reflects important mathematical structures of the underlying system. While the implementation of these methods varies they can be divided into algorithms derived from a discrete Lagrangian \cite{evstatiev_variational_2013,shadwick_variational_2014,stamm_variational_2014,squire_geometric_2012,campos_pinto_variational_2022} and those derived from a Hamiltonian and symplectic formulation \cite{xiao_structure-preserving_2018,kormann_energy-conserving_2021,kraus_gempic_2017}. Many of these methods can significantly reduce numerical heating, improve long-term conservation properties, or relax stability and resolution constraints compared to standard explicit PIC formulations.

To summarize, a diverse range of PIC algorithms is currently available. These methods differ not only in their trade-offs between computational efficiency and numerical accuracy, but also in how particles and fields are coupled, how electromagnetic fields are represented, and which conservation properties are built into the discretization. The diversity of these approaches provides a strong motivation for a framework such as $\pi$-PIC, whose purpose is not to promote one specific algorithm but to facilitate the implementation, testing, and comparison of PIC algorithms with different conservation properties, particle--field coupling strategies, and field representations.

\section{The \texorpdfstring{$\pi$}{pi}-PIC framework}\label{sec:arch}

In recent years, Python has become a common choice for scientific workflows and high-level code development. However, computationally intensive operations are often more efficiently implemented in low-level languages such as C++ or Fortran. To combine usability with performance, $\pi$-PIC employs a Python interface coupled to a precompiled C++ backend, extending the approach introduced in \cite{hi-chi}.

Beyond combining high- and low-level languages, the framework must also accommodate algorithmic developments ranging from runtime diagnostics to modifications of the PIC algorithm itself. To maintain maximum flexibility while maintaining a low barrier to development, the framework introduces three interaction levels: (1) the \textit{Python interface} for simulation setup, runtime control, and diagnostics, (2) \textit{extensions} for local operations on particles and fields, and (3) \textit{solvers} for implementing the particle and field advancement itself.

Through the Python interface, users can initialize fields and particles, configure the PIC algorithm, and advance the simulation state. In addition, predefined particle and field loops provide functionality for accessing and modifying particles and electromagnetic fields during runtime. 

Extensions are modules that perform local actions on electromagnetic fields and particles at every PIC update. Extensions are added to the advance call (describe in \cref{sec:advance}) using the Python interface and are particularly useful for developing new ionization routines or incorporating local collisional and strong-field effects. 

Solvers specify how particles and electromagnetic fields are advanced in time. To ensure compatibility with extensions and the Python interface, solvers follow a predefined interface structure.

The solver and extensions are connected through the advance routine, which manages multi-threading, particle storage and execution order of operations connected to the solver and extensions. In the following sections, we outline these three interaction levels, their functionality and how they are combined in the advance call. 

\subsection{Advance routine}\label{sec:advance}

The advance routine coordinates the execution of solvers and extensions during each PIC advancement step. The main objective of this execution structure is to maximize flexibility in solver development without compromising computational efficiency. 

\begin{figure}[h]
    \centering
    \includegraphics[width=1.0\linewidth]{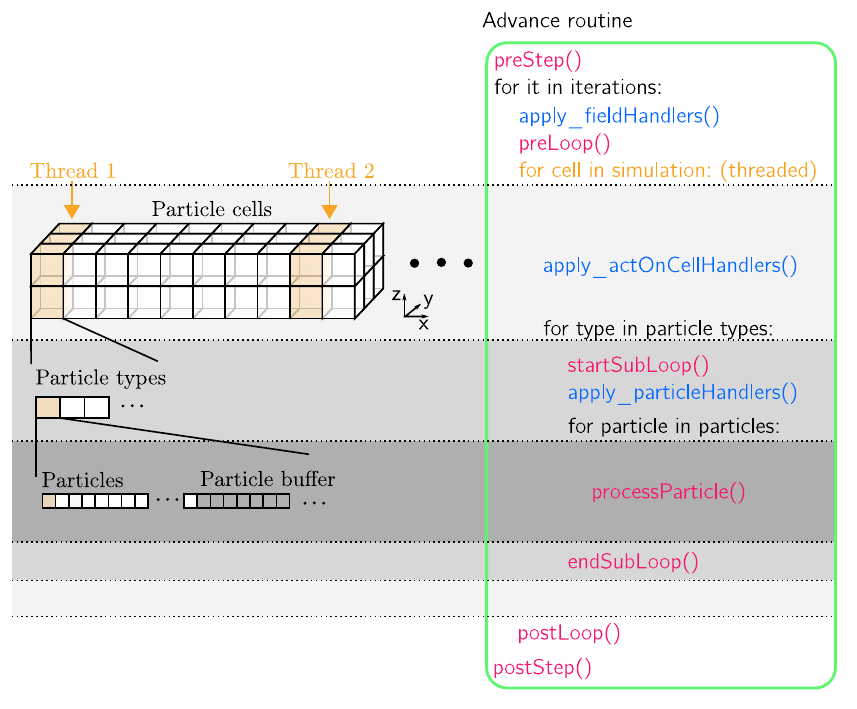}
    \caption{The advance call loops through cells, particle types and particles. During the advance call methods connected to extensions (blue) and solvers (red) are executed. The threading is applied on slices in the $x$-direction with a stride of 8. }
    \label{fig:advance}
\end{figure}

During the advance call, the simulation domain is traversed cell-wise by looping over cells, particle types, and particles (see \cref{fig:advance}). At predefined stages of this traversal, solver methods (red in \cref{fig:advance}) and attached extensions (blue in \cref{fig:advance}) are invoked. The solver is implemented as a C++ class following a predefined interface of mandatory and optional methods, with the red (optional) methods defining the execution order of the PIC update. The blue methods invoke all registered extensions acting on fields, cells, and particles. The advance routine is a method of \texttt{ensemble} which is a structure that manages parallel processing and particle storage in memory.

\begin{figure*}[h]
    \centering
    \includegraphics[width=0.8\textwidth]{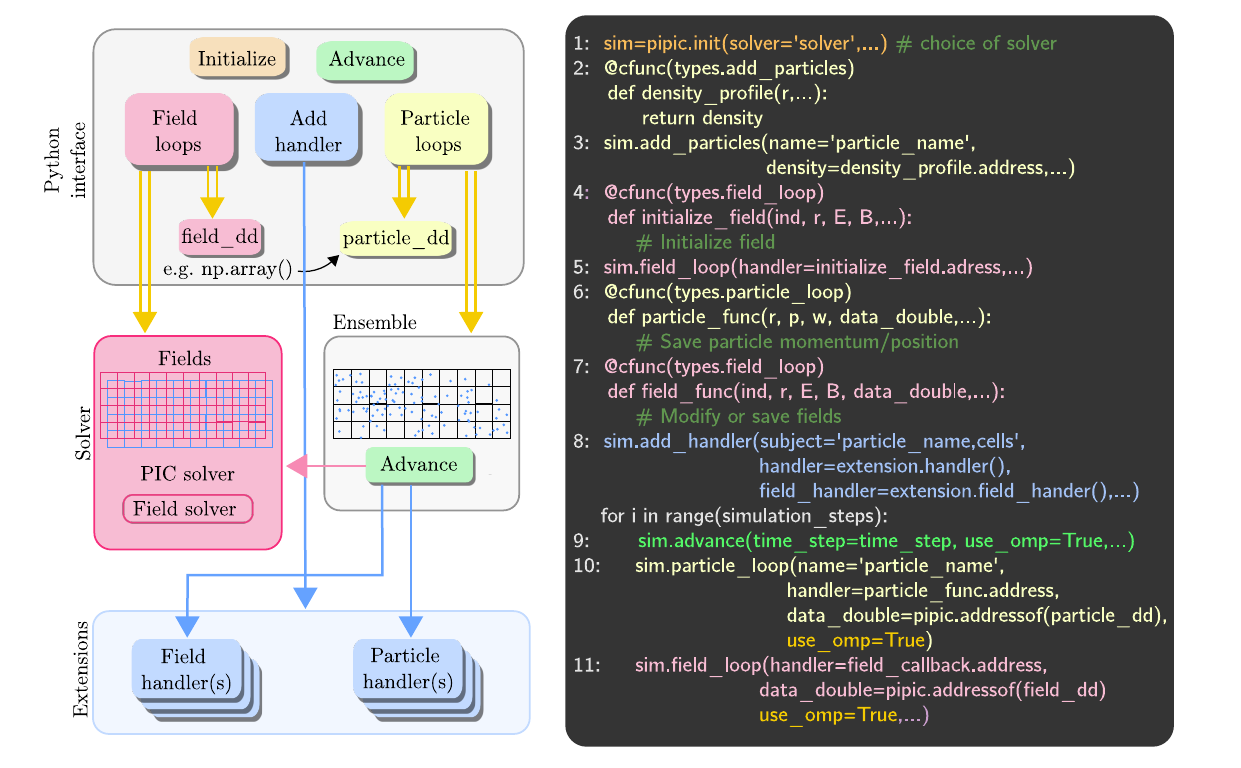}
    \caption{Schematic overview of the code (left) and a simplified main script (right). The different Python commands are color coded to match the activated instances in the overview and \cref{fig:advance}. Arrows symbolize method calls which are activated for the indicated structure. The field and particle loops are applied in a multi-threaded process, indicated with double arrow, if \texttt{use\_omp=True}. For saving the particle and field state an address to the double arrays \texttt{particle\_dd} and \texttt{field\_dd}, respectively, can be passed to the field and particle decorated functions. A complete version of the main script can be found in the Supplementary Material S.1.}
    \label{fig:arc}
\end{figure*}

Threading is applied slice-wise along the x-direction with a stride of 8 cells. To avoid race conditions and prevent particles from being processed multiple times during threaded operations, migrated or newly added particles are temporarily stored in a cell-local particle buffer. Once all threaded operations have completed, the buffered particles are merged into the main particle storage. The particle buffers have an initial capacity of 128 particles per cell and are doubled in size if they are more than half full when particles are merged into main storage.

In summary, the advance routine defines the execution order of the simulation and coordinates the interaction between solvers and extensions (see \cref{fig:advance}). It manages particle storage, migration, buffering, and multi-threading, while ensuring that extensions are executed without race conditions. 


\subsection{Python interface}\label{sec:python_interface}

\begin{table*}
\centering
\caption{Python interface provided by $\pi$-PIC. Functions indicated with (*) require a user-defined decorated function.}
\label{tab:python_interface}
\begin{tabular}{|l|l|l|}
\hline
Class & Method & Purpose \\
\hline
Simulation setup and control & \texttt{init()} & Initialize simulation (and choice of solver) \\
& \texttt{add\_handler()} & Attach extension to advance routine \\
 & \texttt{advance()} & Advance simulation \\
\hline
Particle and field access & *\texttt{add\_particles()} & Add particle density \\
and modification & *\texttt{field\_loop()} & Apply a (threaded) operation to the electromagnetic field \\
& *\texttt{particle\_loop()} & Apply a (threaded) operation to all particles \\
& *\texttt{custom\_field\_loop()} & Sample solver-defined fields (not limited to electromagnetic \\
&  & fields) at predefined coordinates. \\
\hline
Numba decorators & \texttt{add\_particles} & Particle initialization \\
& \texttt{field\_loop} & For implementing field loops\\
& \texttt{particle\_loop} & For implementing particle loops\\
& \texttt{it2r} & For converting index to position \\
& \texttt{custom\_field\_loop} & For implementing custom field loops \\
\hline
\end{tabular}
\end{table*}

A crucial part of a PIC code is the user interface, which should remain intuitive while supporting a broad range of functionality.  In this section, we outline the functionality of the $\pi$-PIC interface by stepping through \cref{fig:arc}, which shows a simplified simulation script together with a schematic overview of the corresponding execution structure. The Python interface methods introduced in this section are summarized in \cref{tab:python_interface}, while the complete example script is included in the Supplementary Material S.1.

In summary, the script initializes the simulation and sets the choice of solver (block 1). Block 2-3 add the particles to the simulation. Block 4-5 set the field. Block 6-7 define particle and field diagnostics. These are later invoked by block 10-11. Block 8 adds predefined extensions to the advance call. In the C++ backend, these extensions are placed in a list and executed in the order they were added. Block 9 advances the simulation, which invokes the advance routine described in the previous section, using multi-threaded processing if \texttt{use\_omp=True}.

The $\pi$-PIC library features three operations for configuring and controlling the particle and field advancement: \texttt{init()}, for initializing the simulation and choosing solver, \texttt{add\_handler()} for adding extensions to the advance routine and \texttt{advance()} for executing the PIC advance, corresponding to block 1, 8 and 9 respectively. In the C++ backend the advance routine is owned by the \texttt{ensemble} structure which also owns and manages the particle storage.

For manipulating and accessing particle states, the library provides \texttt{add\_particles()} and \texttt{particle\_loop()}. The first function adds particles with density defined by the function corresponding to the \texttt{density} argument (that is \texttt{density\_profile()}). The latter provides a handle for looping through all particles and applies the function corresponding to the \texttt{handler} argument (see block 6 and 9). These functions operate directly on the particle data by calling methods of the \texttt{ensemble} structure, and automatically handle the reordering of particles in memory if they are displaced by the particle loop operation.

The electromagnetic field manipulation routine, \texttt{field\_loop()}, follows the same principle as \texttt{particle\_loop()}, namely, it traverses the field data and invokes the function specified by the \texttt{handler} argument (see blocks 7 and 11). Internally, the field loop accesses a method inside the solver for looping through electromagnetic fields.

Operations on particles and fields are exposed through the \texttt{numba.cfunc} decorators listed in \cref{tab:python_interface}. These decorators compile Python functions into C-compatible function pointers, allowing user-defined particle and field operations to be invoked directly from the underlying C++ code. Provided that the function signature matches the arguments specified by the decorator and variables defined outside the function are not modified within it, arbitrary operations can be implemented. This approach combines the flexibility of Python with the performance benefits of precompiled code.

The limitation that prevents changing Python objects outside decorated functions arises because the decorator turns the function into standalone C code that executes independently of the Python runtime. Consequently, Python objects outside the function cannot be modified directly, and are treated as compile-time constants. Dynamic data exchange must occur through an alternative C-compatible interface. Therefore, the last arguments of every signature are a \texttt{data\_double} and \texttt{data\_int}, which corresponds to static C pointer referencing a float64 and an int32 array. The arrays may be used to store, retrieve, or communicate data between successive PIC update steps (see block 10 and 11), providing a mechanism for accessing field and particle states for diagnostics. Note that, when multi-threading is enabled for the field or particle loops, different threads must not write to the same memory location, as this results in undefined behavior. However, if \texttt{data\_double} or \texttt{data\_int} points to an array, different threads may safely write to different elements of that array. An example of this de-referencing procedure is provided in Supplementary Material S.1.

More advanced field diagnostics can be implemented using \texttt{custom\_field\_loop()} together with the \texttt{it2r} and \texttt{custom\_field\_loop} decorators. The function associated with the \texttt{it2r} decorator defines a mapping from a set of indices to positions in the simulation domain, while the function associated with the \texttt{custom\_field\_loop} decorator transfers the corresponding field values to a user-defined array. We shall return to this functionality in \cref{sec:diagnostics}, since it can be used to expose solver-specific field quantities such as scalar and vector potentials. The exact behavior of \texttt{custom\_field\_loop()} is therefore solver dependent, allowing each solver to determine which field variables are accessible and how they are exposed through the Python interface. For a complete implementation example, see the example script in Supplementary Material S.1.

In summary, the Python interface has functionality to initialize the simulation, field and particle state, import and attach extensions to the advance call, modify and read particle and field state as well as advance the state of the simulation. Although the interface itself is functionally simple, it still offers enough capabilities to support a wide range of diagnostics, runtime controls, and localized operations on both particles and fields. For example, particles may be removed by setting their weight to zero, fields may be modified or damped, simulations may be terminated once specified conditions are met, and diagnostics can be tailored to the requirements of a particular study.

The design philosophy of $\pi$-PIC is to keep the Python interface focused on simulation setup, control, and diagnostics, while more specialized or performance-critical functionality is implemented through extensions. Consequently, extensions provide the preferred mechanism for developing reusable particle and field operations, as discussed in the next section.

\subsection{Extensions}\label{sec:extensions}

Extensions are compiled Python modules for local modification of fields and particles. They traverse the field and/or particle data and apply user-defined operations at each field point, cell, or particle. The extensions are written in C++ and imported into Python using pybind11. There are two extension methods \texttt{field\_handler()} and \texttt{handler()}. The first provides functionality for modifying electromagnetic fields, and the second for adding particles and changing particle state. The field modification is computationally the same as for the field loop in the Python interface, while the particle handling is applied in conjunction with the particle processing defined by the solver in a multi-threaded manner (see \cref{fig:advance}). 

\begin{table*}[h]
\centering
\caption{Solver interface methods provided by $\pi$-PIC. Methods indicated with (*) are mandatory.}
\label{tab:solver_methods}
\begin{tabular}{|l|l|l|}
\hline
Class & Method & Purpose \\
\hline
\texttt{pic\_solver} & \texttt{*advance()} & Mandatory PIC advancement call \\
& \texttt{preStep(timeStep)} & Pre-advance operations \\
& \texttt{preLoop()} & Operations before particle traversal \\
& \texttt{startSubLoop()} & Operations at start of cell traversal \\
& \texttt{processParticle()} & Threaded particle processing \\
& \texttt{endSubLoop()} & Operations at end of cell traversal \\
& \texttt{postLoop()} & Operations after particle traversal \\
& \texttt{postStep(timeStep)} & Post-timestep operations \\
\hline
\texttt{field\_solver} & \texttt{*fieldLoop()} & Access for interface and extensions \\
& \texttt{*cellSetField()} & Transfer field data to \texttt{cellInterface} \\
& *\texttt{customFieldLoop()} & Custom access to field data \\
& *\texttt{advance(timeStep)} & Field advancement \\ 
\hline
\end{tabular}
\end{table*}

To ensure compatibility with the rest of the system, the arguments of \texttt{field\_handler()} and \texttt{handler()} are predefined. For \texttt{field\_handler()} these consist of pointers to grid index, position and the electromagnetic field. For \texttt{handler()}, the arguments consist of raw pointers to particle data in the current cell. The translation between raw pointers and C++ objects is handled by the \texttt{cellInterface} structure, which provides a C++ abstraction of the underlying pointer-based data.

Although extensions are typically implemented in C++, the raw-pointer interface between the advance routine and extensions also enables implementations in other languages. For \texttt{add\_handler()} in the Python interface, the only requirement is a compiled function handle whose input arguments correspond to raw pointers matching the expected interface. This enables extensions written in Fortran, Rust, and other callable languages. However, incorporating extensions written in these languages is more complicated, since the \texttt{cellInterface} must first be decoded from the pointer array corresponding to the \texttt{cellInterface} structure. To demonstrate this example extensions using Rust and Fortran are available in \cite{hi-chi2023pipic}.

Internally, particles are organized by cell and particle type according to the execution order shown in \cref{fig:advance}. Three extension methods are invoked within the advance call: \texttt{apply\_fieldHandlers()}, \texttt{apply\_actOnCellHandlers()}, and \texttt{apply\_particleHandlers()}. In turn, these invoke the \texttt{field\_handler()} or \texttt{handler()} functions registered through \texttt{add\_handler()} (see \cref{fig:arc} block 8). The \texttt{field\_handler()} is executed for all field points, while \texttt{handler()} is executed either once per cell or once per particle, depending on whether the \texttt{subject} argument in the Python \texttt{add\_handler()} call is set to \texttt{cells} or to a particle type. Multiple registered handlers are executed in the order in which they were added. Note that, in accordance with the advance scheme, all \texttt{field\_handler}s are executed before any particle or cell \texttt{handler}s

The execution model imposes some restrictions on extension design. Particle insertion must occur during the execution of \texttt{apply\_cellHandler()}; otherwise newly created particles are placed in the particle buffer together with migrated particles and will not be processed until the next timestep. Furthermore, while both particles and fields may be modified during the same timestep, they cannot be modified simultaneously within the same extension. As a result, while simple local particle and field operations may be implemented as extensions, more sophisticated models -- such as collisional methods involving field corrections, conservation enforcement, or distribution-function representations -- are generally better suited for implementation at the solver level.

\subsection{Solvers}\label{sec:solver_arc}

In $\pi$-PIC, solvers define the particle and field update. A solver consists of two core classes: a \texttt{pic\_solver} and a \texttt{field\_solver}, which contains the particle and field update. The \texttt{field\_solver} is a mandatory member of the \texttt{pic\_solver} class, reflecting the fact that field and particle updates are generally coupled and must be advanced consistently. Both classes follow a predefined interface consisting of the methods summarized in \cref{tab:solver_methods}. 

To maintain compatibility with the rest of the framework, the \texttt{field\_solver} must implement two methods: \texttt{fieldLoop()} and \texttt{cellSetField()}. These methods provide a translation layer between the solver's internal field representation and the Cartesian electric and magnetic field representation used by extensions. The \texttt{fieldLoop()} method is responsible for accessing and updating field values, and is accessed both by the Python interface in the \texttt{field\_loop()} call and by the extensions by the \texttt{field\_handler()}. The \texttt{cellSetField()} method is invoked by the advance routine to populate the \texttt{cellInterface} with field data that can subsequently be accessed by particle extensions. This design allows different types of electromagnetic grids and field representations to be implemented. However, it requires developers to provide methods that translate the internal field representation to the Cartesian grid used by the framework, as well as methods for updating the solver's internal field representation through this interface.

Note that, compatibility between solvers and extensions should be understood as execution compatibility rather than physical consistency. Extensions developed for one solver will generally execute with another solver provided they satisfy the interface contract. However, conservation properties and the physical validity of the resulting combined algorithm are not guaranteed by the framework and must be verified by the developer.

The \texttt{pic\_solver} class defines a single mandatory method, \texttt{advance()}, which implements the PIC update. Although this method can be implemented independently of the rest of the framework, the recommended approach is to base it on the \texttt{advance()} method presented in \cref{sec:advance}, as this provides built-in multi-threading, extension compatibility, and particle management. The method \texttt{ensemble::advance()} invokes the optional methods of \texttt{pic\_solver}: \texttt{preStep()}, \texttt{preLoop()}, \texttt{startSubLoop()}, \texttt{processParticle()}, \texttt{endSubLoop()}, \texttt{postLoop()} and \texttt{postStep()} and, thereby, defines the execution order as described in \cref{fig:advance}. The intended use is to distribute the particle and field update operations among these methods in a manner appropriate for the numerical scheme being implemented. Note that, particles migrating between cells are redistributed after completion of the multi-threaded particle loop (see \cref{fig:advance}). Consequently, particles moved in \texttt{preStep()}, \texttt{preLoop()}, \texttt{postLoop()}, or \texttt{postStep()} may not be reassigned to their new cells until the next particle loop. This can be avoided by invoking the \texttt{ensemble} method \texttt{particleLoop()}, which uses the same multi-threaded implementation as the Python function \texttt{particle\_loop()} to redistribute particles. 

In summary, the solver interface separates the framework infrastructure from the numerical implementation of the PIC algorithm. The \texttt{ensemble} is responsible for simulation management, particle storage, and extension handling, whereas the solver defines the particle update, field update, and their coupling. This separation allows diverse numerical formulations to be incorporated while remaining compatible with the surrounding framework.

\subsection{Supported solver architectures and limitations}\label{sec:solvers_comp}

The class structure of \texttt{pic\_solver} allows for a large degree of freedom in terms of the implementation of the PIC update, for example, solvers can own arbitrary field structures (e.g., collocated electromagnetic fields, staggered Yee grids, vector-potentials and spectral field representations) and particle operations (e.g., shape functions, deposition schemes, particle-field corrections and field gathering). In the following section, we discuss which classes of PIC algorithms can be represented within this architecture and identify its current limitations.

\subsubsection{Particle--field coupling}\label{sec:field_particle_coupling}
The solver owns the particle update, field update, particle--field coupling, shape functions, field gathering, and particle deposition. These operations are not prescribed by the framework and may therefore be implemented according to the requirements of the numerical method. This allows both conventional explicit PIC schemes, featuring decoupled particle push, deposition, field advance and field gathering, as well as methods employing coupled particle--field updates, such as implicit methods, energy-conserving schemes, and certain classes of variational and structure-preserving methods. Similarly, collisional operators may be implemented directly within the solver when they form an integral part of the numerical update. 

This design choice reflects the wide variety of coupled particle--field algorithms. Implicit PIC methods, variational formulations, and finite-element based approaches often require substantially different execution patterns, data structures, and nonlinear solution strategies. Consequently, the framework currently provides the basic execution infrastructure while leaving the implementation of solver-specific parallelization strategies to the developer.

To demonstrate that coupled particle--field updates can be accommodated within the framework, we present an implementation of a one-dimensional implicit electrostatic solver in \cref{sec:implicit}.

\subsubsection{Parallelization and particle management}\label{sec:parallelization}

Parallelization and particle memory management are handled automatically at both the Python interface and extension levels. In the Python interface, decorated field and particle loops are executed in parallel when \texttt{use\_omp=True}. If particles are displaced during \texttt{particle\_loop()}, they are automatically redistributed to their new cells once the loop has completed.

Similarly, extensions are executed through the framework's multi-threaded advance routine without requiring explicit parallelization by the developer. Particle migration is handled automatically by the framework: newly added particles are redistributed immediately after \texttt{apply\_actOnCellHandlers()}, allowing them to participate in the remaining stages of the advance loop, while particles that migrate between cells during particle processing are redistributed after completion of the advance routine. However, when writing extensions, care must be taken when using thread-shared resources. For example, deterministic random-number generation requires a thread-aware implementation. 

At the solver level, the framework provides built-in parallelization for the methods \texttt{startSubLoop()}, \texttt{processParticle()}, and \texttt{endSubLoop()}, which are executed as part of the multi-threaded particle traversal in the advance routine. Solver methods outside this execution path, such as \texttt{preStep()}, \texttt{preLoop()}, \texttt{postLoop()}, and \texttt{postStep()}, may likewise operate on fields and particles. However, since these operations are solver defined, the framework does not prescribe a parallel execution strategy, leaving the implementation of any required parallelization to the solver developer.

Similarly, if particles are modified outside \texttt{processParticle()}, any required synchronization and particle management remain the responsibility of the solver developer. Particle redistribution is, however, facilitated by the \texttt{ensemble::particleLoop()} method, which redistributes particles according to their new cell assignment using the framework's built-in multi-threaded implementation. Likewise, the implementation and parallelization of \texttt{fieldLoop()} are entirely solver defined.

\begin{table*}[!b]
\centering
\caption{Built-in extensions currently available in $\pi$-PIC. (*) indicates extensions described in this article.}
\label{tab:extensions}
\begin{tabular}{|l|l|}
\hline
Extension & Purpose \\
\hline

\multicolumn{2}{|c|}{Strong field quantum electrodynamics} \\
\hline
\texttt{qed\_gonoskov2015}\cite{gonoskov.pre.2015} & QED event generator based on rejection sampling and subcycling \\
\texttt{qed\_volokitin2023}\cite{volokitin.jcs.2023} & QED event generator with an optimized rate computation \\
*\texttt{focused\_pulse} & Tight focusing of laser pulses\\
\texttt{downsampler\_gonoskov2022}\cite{gonoskov.cpc.2022} & Conservative particle downsampling preserving  moments and distributions \\
\texttt{landau\_lifshitz} & Radiation-reaction force using the Landau--Lifshitz model \\
\hline
\multicolumn{2}{|c|}{Boundary handling} \\
\hline
*\texttt{absorbing\_boundaries} & Masking implementations of absorbing boundary for fields and particles \\
*\texttt{moving\_window} & Moving-window implementation with particle injection \\
 \texttt{x\_reflector\_c} & Reflecting boundaries for electromagnetic fields \\
 \texttt{x\_converter\_c} & Field conversion utility for diagnostics and boundary treatments \\
\hline

\end{tabular}
\end{table*}

\subsubsection{Field representations}\label{sec:field_representation}
The framework places no restrictions on the internal field representation. Solvers may employ collocated electromagnetic fields, spectral field representations, vector-potential formulations, generalized momentum formulations, or other solver-defined field structures. However, to maintain compatibility with extensions and the Python interface, the solver must provide implementations of \texttt{fieldLoop()} and \texttt{cellSetField()}. These methods define the interface between the solver's internal field representation and the Cartesian electric and magnetic field representation used by the framework, allowing data to be transferred in both directions. This translation layer is used exclusively by extensions and the Python interface and does not restrict the internal field representation used by the solver.

\subsubsection{Mesh topology}\label{sec:mesh_topology}
The current limitation of the framework lies primarily in the particle infrastructure rather than in the field representation. While field data may be represented using arbitrary data structures, particle storage, migration, and buffering are currently based on the Cartesian \texttt{cellInterface} structure. Consequently, particle handling on unstructured or finite-element meshes has not yet been demonstrated within the framework. In principle, the \texttt{cellInterface} and associated initialization routines could be generalized to alternative mesh topologies. However, such an addition to the framework would require modifications to particle initialization, migration, and parts of the advance routine and is therefore left for future work.

\subsubsection{Diagnostics}\label{sec:diagnostics}
The Python interface currently exposes electric and magnetic fields through the standard diagnostic routines. Alternative field variables, such as scalar potentials, vector potentials, generalized momenta, gauge quantities, or solver-specific auxiliary fields, may be exposed to the Python interface through the implementation of \texttt{customFieldLoop()} which links to the Python interface function \texttt{custom\_field\_loop()}. Note that, the implementation of this function is solver defined, allowing each solver to determine which field variables are exposed and how they are represented.

\subsubsection{Current limitations and future developments}\label{sec:conclusion_solver_comp}
In summary, $\pi$-PIC imposes relatively few restrictions on the numerical algorithm itself. The principal limitations arise from the particle infrastructure, which is presently based on a structured Cartesian cell decomposition. To support non-cartesian mesh topologies, the \texttt{cellInterface} could be generalized to represent alternative grid structures, such as spherical or unstructured meshes. 

In addition, the built-in multi-threading and particle-management routines are optimized for conventional particle processing. To facilitate the implementation of solvers with stronger particle–field coupling, as well as operations commonly employed in collisional PIC methods, the framework would benefit from additional predefined functionality, such as binary particle search routines and alternative data-access patterns. These aspects represent natural directions for future development of the framework.

\section{Implemented extensions}\label{sec:appl} 
Beyond the generic extension interface, $\pi$-PIC currently provides a collection of extensions implementing common operations used in plasma simulations. These include strong-field QED modules, particle downsampling utilities, field-manipulation routines, and boundary-condition implementations. The currently available extensions are summarized in \cref{tab:extensions}. Together, they demonstrate how additional physics and numerical functionality can be incorporated independently of the underlying solver.

In this section, we present two extensions for handling boundaries: an implementation of absorbing boundaries with replacement of particles (\cref{sec:absbound}) and a moving window (\cref{sec:movingwin}). Additionally, we present an extension for mapping a spatially extended pulse to its focus (\cref{sec:tightfocus}). 

\subsection{Absorbing boundary layer}\label{sec:absbound}

Many plasma systems are most naturally modeled using open boundaries, corresponding to a plasma that extends beyond the computational domain. Among the most widely used emulators for open boundaries in PIC codes are perfectly matched layers (PMLs) \cite{berenger_perfectly_1994}, which employ complex coordinate stretching to mimic an absorbing region within a finite number of grid cells. 

An alternative approach, used primarily in MHD and two-fluid plasma simulations, is provided by lacunae-based boundary conditions \cite{meier_modeling_2012}, which uses an auxiliary exterior-domain construction that approximates wave propagation in an unbounded domain. 

For PIC simulations, open boundaries involve additional complexity since particles must be removed and reintroduced in a manner that avoids the generation of spurious electromagnetic fields \cite{lehe_absorption_2022}. While a variety of approaches have been proposed (e.g., particle re-injection \cite{klimas_particle--cell_2008, miller_extended_2021} and PML-based methods \cite{lehe_absorption_2022}), no universally applicable treatment has emerged, and the appropriate boundary model often depends on the physical system and numerical method under consideration.

Here, we adopt a simple masking (or sponge-layer) approach, in which electromagnetic fields are gradually attenuated within a boundary region \cite{tajima_absorbing_1981}. Particles are removed in the same region and replaced according to a prescribed thermal distribution. Compared with PML and lacunae-based methods, the masking approach is considerably simpler to implement, as it often can be implemented largely independently of the underlying solver. This simplicity comes at the expense of a less sophisticated treatment of outgoing waves and particles. In particular, the absorption efficiency depends strongly on the angle at which waves impinge on the boundary, as demonstrated later.

The primary purpose of this section is to demonstrate the utility of extensions in $\pi$-PIC. Therefore, we do not attempt a systematic comparison with existing open-boundary methods, but instead focus on demonstrating the implementation and benchmarking capabilities of the framework. It should also be noted that extensions provide local operations on fields and particles independent of the underlying solver. Consequently, more sophisticated boundary treatments, particularly those tied to the underlying discretization and those requiring global access to the particle or field state, may be better suited for implementation within the solver itself.

\begin{figure}[t]
    \centering
    \includegraphics{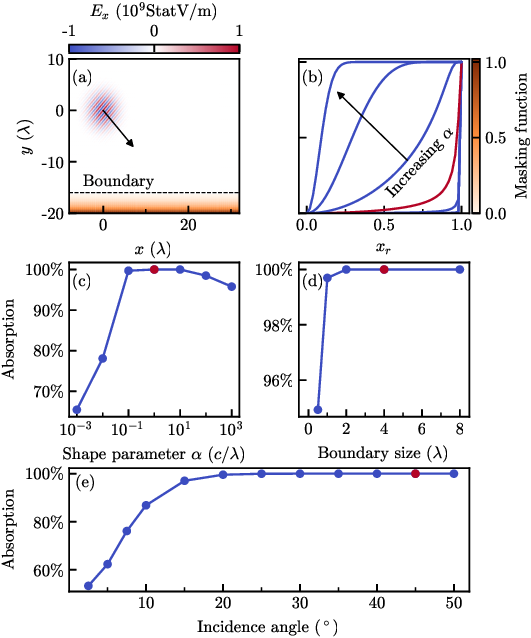}
    \caption{Panel (a) shows the initial pulse propagating towards the boundary at an incidence angle of $45^\circ$, with the masking function $R(y)$ shown in orange. Panel (b) shows $R(y)$ for different values of the shape parameter $\alpha$. The red curve corresponds to $\alpha c/\lambda = 1$, which is also indicated by the red marker in panel (c). Panel (c) shows the resulting absorption as a function of the shape parameter $\alpha$, while panels (d) and (e) show the dependence of the absorption on boundary width and incidence angle, respectively. In panels (c)-(e), the red markers indicate the default values of the remaining parameters used while varying the parameter shown on the horizontal axis of panel (c)-(e). The absorption was calculated as the decrease in total electromagnetic energy before and after interaction with the boundary.}
    \label{fig:bc}
\end{figure}

The masking method is implemented by attenuating the electromagnetic fields at every timestep according to

\begin{equation*}
\tilde{E}(x) = R(x)E(x), \quad \tilde{B}(x) = R(x)B(x),
\end{equation*}
where $R(x)$ is the masking function and $\tilde{E},\tilde{B}$ are the updated fields. To avoid absorption changing with the length of timestep, we require that the cumulative attenuation after $n$ timesteps $\Delta t$, to be equal to the attenuation obtained using a single timestep of size $n\Delta t$, that is $R(x,\Delta t)^{n}=R(x,n\Delta t)$. This condition is fulfilled by the choice 
\begin{equation*}
    R(x,\Delta t)=e^{\Delta t r(x)}, 
\end{equation*}
where $r(x)$ is a function chosen such that reflection is minimized at the boundary, that is $r(x_{b})=0$ and $\frac{dr(x_{b})}{dx}=0$, where $x_b$ is the start of the boundary. Additionally we require that $\lim_{x\rightarrow x_{\text{max}}}R(x) \rightarrow 0$, where $x_{\text{max}}$ is the end of the boundary. This gives the condition $\lim_{x\rightarrow x_{\text{max}}}r(x)\rightarrow-\infty$. A function that satisfies these conditions is 
\begin{equation*}
    r(x_r) = \alpha\left[\cos(\pi x_r/2)-\frac{1}{\cos(\pi x_r/2)}\right],
\end{equation*}
where $x_r = \frac{x-x_b}{x_{\text{max}}-x_b}$ is the relative, normalized position at the boundary such that $x_r\in[0,1]$ and $\alpha$ is a shape parameter controlling the steepness of the absorption, see \cref{fig:bc}.(b). To keep $\alpha\Delta t $ close to unity and the exponent dimensionless, $\alpha$ is scaled by $\lambda/c$, where $\lambda$ is the typical wavelength of the radiation and $c$ the speed of light.

We assess the absorption by letting a Gaussian beam with wavelength $\lambda=\SI{1}{\mu\meter}$, pulse duration $t_s=\SI{10}{\femto\second}$ and equal transverse extent, propagate towards the boundary as shown in \cref{fig:bc}. The simulation had a space step of $\Delta x = \lambda / 8$ and timestep $\Delta t = \Delta x / 4c$. From the result shown in \cref{fig:bc}, we find acceptable absorption for $\alpha\in[0.1,10]$ using a boundary size of $4\lambda$. Moreover, \cref{fig:bc}(e) shows that the absorption remains relatively insensitive to the angle of incidence for angles above $20^\circ$, but decreases rapidly below this threshold.

To summarize: while the extension effectively attenuates outgoing waves, the absorption depends on the angle of incidence. We consider the extension a useful demonstration of how absorbing boundaries can be implemented within the framework. The implementation is available in \cite{hi-chi2023pipic}. 

\subsection{Moving window}\label{sec:movingwin}
For simulations involving, for example, laser-plasma interaction the computational demands can typically be reduced by only simulating a region of space that moves with the process of interest, using a so-called \textit{moving window}.

The available solvers in $\pi$-PIC are all implemented with spectral boundary conditions, consequently a propagating field will reappear at the opposite side of the simulation box if it passes through the boundary. This makes the implementation of the moving window computationally efficient: for a window moving with some speed, fields and plasma perturbations are simply removed within a region that moves around the simulation box at the same velocity. The full implementation and use of the moving window is documented in \cite{hi-chi2023pipic}.

\subsection{Mapping of tightly focused laser pulses}\label{sec:tightfocus}

 \begin{figure*}
    \centering
    \includegraphics{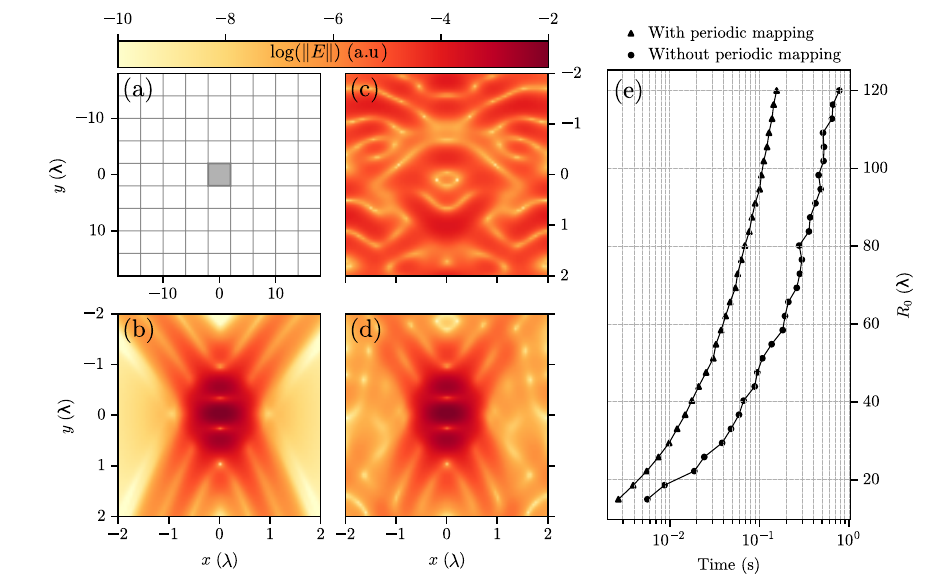}
    \caption{Panel (a) shows the full simulation region containing the initial pulse with radius $R_0$, (yellow-red colormap), and its focal region (shaded gray). By applying the periodic mapping, the pulse shown in (a) can be mapped into the shaded region, the resulting field configuration is presented in panel (c). Advancing the pulse to its focus, corresponding to advancing one timestep $R_0/c$ using a spectral solver, produces the field shown in panel (b) for the simulation of the full domain of (a) and in panel (d) where the periodic mapping has been used. Panel (e) compares the average runtime across $15$ runs for the initialization and propagation of the electromagnetic field using the two methods: triangles denote focusing with periodic mapping and dots without.}
    \label{fig:periodic_mapping_case}
\end{figure*}

Tightly focused laser pulses are central to studies of high-intensity laser–matter interactions \cite{Mourou2006, Marklund2006}, fundamental quantum systems \cite{DiPiazza2012} and various areas of attosecond physics \cite{Krausz2009}. While analytical methods exist \cite{Davis1979, Couture1981, Fedotov2007, Sepke2006, Salamin2006} to describe the propagation of their associated electromagnetic fields, numerical approaches provide a more direct and flexible alternative, free from the inaccuracies introduced by simplifying assumptions. Such computations are particularly relevant for vacuum electron acceleration \cite{Popov2008}, and radiation generation driven by laser–electron dynamics, where interactions are highly sensitive to the structure of the focal electromagnetic field \cite{Cardenas2019, Gonoskov2011, Harvey2016}. However, numerical techniques based on direct integration of Green’s functions or FDTD methods can be computationally intensive and prone to numerical dispersion. Spectral field solvers on the other hand, as implemented in $\pi$-PIC, leverage fast Fourier transforms to evolve the electromagnetic field. This enables dispersion-free computation of the focal field of tightly focused laser pulses, given a known far-field profile, within a single timestep. Despite this advantage, the simulation box must be sufficiently large  to encompass the initial laser pulse profile. Consequently, excessive computational effort is spent evolving the simulation in regions where no significant dynamics occur, even though the physical interaction of interest is many times confined to a much smaller box (see the shaded region in \cref{fig:periodic_mapping_case}.(a)).

To avoid this computational overhead, the periodic boundary conditions imposed by spectral solvers can be leveraged, as is demonstrated in \cite{panova2021optimized}. In that work, periodicity in the longitudinal direction was used to reduce the computational domain. Here, we extend this approach by exploiting periodicity in all three spatial dimensions. We further introduce an optimized algorithm for selecting the remapping region. The linearity of Maxwell's equations imposes that electromagnetic signals propagate independently. As such, it is possible to map a large focusing pulse to a smaller periodic space, and let it propagate to focus without affecting its evolution. Alternatively, this can be understood as the time-reversed evolution of a pulse in a periodic domain. As the pulse defocuses, it crosses the boundary and reenters from the opposite side, interfering with the field in the region. Reversing this process corresponds to a pulse focusing from a periodic domain into a localized pulse.

\begin{table*}[!b]
\centering
\caption{Built-in solvers currently available in $\pi$-PIC.}
\label{tab:solvers}
\begin{tabular}{|l|l|l|l|}
\hline
Name & Field solver & Particle update & Purpose \\
\hline
\texttt{fourier\_boris}
& Spectral
& Boris pusher
& Reference electromagnetic PIC solver \\
\hline
\texttt{ec}
& Spectral
& 1st order energy-conserving scheme\cite{gonoskov_jcp_2024}
& Explicit energy-conserving PIC \\
\hline
\texttt{ec2}
& Spectral
& 2nd order energy-conserving scheme\cite{gonoskov_jcp_2024}
& Higher-order energy-conserving PIC \\
\hline
\texttt{electrostatic\_1d}
& \multicolumn{2}{l|}{Coupled electrostatic particle--field update\cite{markidis_energy_2011}} & Reference solver for \\
\texttt{\_implicit} &  \multicolumn{2}{l|}{} & development and testing  \\
\hline
\texttt{electrostatic\_1d}
& Electrostatic 
& Leapfrog particle pusher
& Same as above\\
\hline
\end{tabular}
\end{table*}

Consequently, the mapped electric (and magnetic) field $\vec{E'}$ can be written as a superposition of the field in a space fully enclosing the initial pulse
\begin{equation*}
\vec{E'}\left[i, j, k \right]
= \sum_{h_x,h_y,h_z} \vec{E} \left[h_x n_x + i, h_y n_y + j, h_z n_z + k \right],
\end{equation*}
where  $i,j,k$ denote grid indices, $n_x, n_y, n_z$ are the number of cells along each coordinate axes and the integers $h_m$ with $m \in \{x,y,z\}$ are referred to as \textit{hypercells}, which are periodic replicas of the simulation box arranged to fully enclose the initial pulse profile.
The limits of $h_m$ can be extracted by assuming that the laser pulse initially occupies a spherical segment with thickness $\ell$, aperture angle $\theta_\text{max}$, and radius $R_0$. The latter is interpreted as the distance from the pulse centroid to the location of the focus. The hypercell bound for each coordinate axis is then given by
\begin{equation*}
h_m^\pm = \left\lfloor \frac{L_m/2 \pm (R_0 + \ell/2)}{L_m} \right\rfloor,
\end{equation*}
where $L_m$ is the length of the simulation box. An example of the simulation box is depicted in \cref{fig:periodic_mapping_case}.(a) as the shaded quadratic region, whereas the three bottom rows of unshaded squares demarcate the required hypercells to accurately capture the pulse profile. To avoid superimposing negligible electromagnetic fields, arising from empty cells, only the coordinates $\vec{r}$ that satisfy 
\begin{equation*}
\left| R_0 - r \right| \leq \frac{\ell}{2}, \quad \theta = \arccos(\hat{r} \cdot \hat{n}) \leq \theta_{\text{max}},
\end{equation*}
are included. Here, $r = |\vec{r}|$, $\hat{r} = \vec{r}/r$, and $\hat{n}$ is the unit vector along the direction of $R_0$. The specific implementation of the periodic mapping, including an optimized procedure, is provided in \cite{hi-chi2023pipic}.

To demonstrate the periodic mapping, we performed 3D simulations for a linearly polarized focused pulse with far‐field pulse profile \(u\) analogous to that given in \cite{panova2021optimized}:
\begin{equation*} 
    u(\vec{r}) = \frac{u_{\|}(r-R_0)}{r}\, u_{\perp}(\theta),
\end{equation*}
where
\begin{equation*} 
    u_{\|}(\xi) = \sin \left(\frac{2\pi \xi}{\lambda}\right) \cos^2 \left(\frac{\pi \xi}{\ell}\right) \, \Pi \left(\xi, -\frac{\ell}{2}, \frac{\ell}{2}\right),
\end{equation*}
is the longitudinal profile and \(\Pi \left(\xi, \xi_{\text{min}}, \xi_{\text{max}}\right)\) is the rectangular function that equals unity whenever \(\xi \in [\xi_{\text{min}}, \xi_{\text{max}}]\) and zero otherwise. The transverse shape is given by
\begin{gather} 
u_{\perp}(\theta) = \Pi \left(\theta, 0, \theta_{\text{max}} - \frac{\varepsilon}{2}\right) + \nonumber \\
+\cos^2 \left(\frac{\pi\left(\theta - \theta_{\text{max}}+\frac{\varepsilon}{2}\right)}{2 \varepsilon}\right) \, \Pi \left(\theta, \theta_{\text{max}}-\frac{\varepsilon}{2}, \theta_{\text{max}}+\frac{\varepsilon}{2}\right), \label{eq:transverse_shape}
\end{gather}
where $\varepsilon = 0.1$ is an angle that allows for a smooth transition of \cref{eq:transverse_shape} around $\theta_{\text{max}}$. The pulse was initialized using \(\theta_{\text{max}} = 59^\circ\), \(\ell = 2\lambda\), \(R_0 = 16\lambda\), and advanced within a quadratic box with side lengths $2(R_0 + \ell)$, which corresponded to the full extent of \cref{fig:periodic_mapping_case}.(a). Additionally, a separate pulse was initialized using the periodic mapping implementation within a smaller domain with side lengths $4\lambda$, corresponding to the shaded region in \cref{fig:periodic_mapping_case}.(a). Both configurations had a spatial resolution of $16$ cells per wavelength and were evolved over a single timestep $R_0/c$, yielding the final electromagnetic field states as shown in \cref{fig:periodic_mapping_case}.(b,d). The relative root-mean-square error between the two configurations for the electromagnetic energy density computed within the focal region, was found to be \(\SI{1.43e-4}{}\), demonstrating the accuracy of the periodic mapping implementation. 

To assess the efficiency of periodic mapping, we compared the runtime of 2D simulations between the two configurations as a function of the propagation length $R_0$, using Intel(R) Core(TM) i7-10700K CPU with 8 physical cores and 16 threads. The opening angle and spatial resolution were kept fixed throughout. The pulse was initialized and propagated to focus in a single timestep of duration $R_0/c$. For simulations without periodic mapping, the computational domain was chosen to fully enclose both the focal region and the pulse profile, such that \(L_x = R_0+\ell/2+2\lambda\) and \(L_y = 2R_0 \sin \left(\varepsilon + \theta_{\text{max}}\right)\). 

The results of this benchmark, presented in \cref{fig:periodic_mapping_case}.(e), demonstrates a significant speed-up for the periodic mapping algorithm across all values of \(R_0\).

\section{Solver benchmark examples}\label{sec:solvers}

$\pi$-PIC includes a collection of electromagnetic and electrostatic solvers intended both for production simulations and as examples of how new numerical methods can be integrated into the framework. The available solvers are summarized in \cref{tab:solvers}. $\pi$-PIC features three electromagnetic solvers. These consist of a standard Fourier-Boris (FB) solver, combining a spectral Maxwell solver with a Boris particle pusher, and two variants of the energy-conserving solver. The latter has previously been benchmarked within the $\pi$-PIC framework using scalings of Landau damping, and resolution-convergence studies for relativistic laser--solid reflection \cite{gonoskov_jcp_2024}. In addition, $\pi$-PIC provides two one-dimensional electrostatic solvers: one explicit and one implicit. These solvers are primarily intended as minimal examples illustrating how new numerical methods can be incorporated into the framework. 

In the following sections, we introduce the electrostatic solvers and compare them using a two-stream instability benchmark (\cref{sec:two_stream_benchmark}). Subsequently, we present laser wakefield acceleration simulations (\cref{sec:benchmark}) carried out with the Fourier-Boris and the energy-conserving solver (EC). As a reference, these results are compared to corresponding simulations performed with the PIC code Smilei \cite{derouillat_smilei_2018} employing the M4 solver \cite{lu_time-step_2020}. Together, these benchmarks demonstrate the framework's ability to accommodate different PIC formulations and facilitate the comparison of numerical solvers.

\subsection{Implicit and explicit electrostatic solvers}\label{sec:two_stream_benchmark}
To demonstrate the implementation of solvers, two one-dimensional electrostatic solvers were implemented in $\pi$-PIC. The first solver is an explicit scheme, presented in the next \cref{sec:explicit}, while the second is an implicit scheme, discussed in \cref{sec:implicit}, based on the implementation outlined in \cite{markidis_energy_2011}. The source code for both solvers is provided as Supplementary Material S.2 and S.3.

\subsubsection{An explicit one-dimensional electrostatic solver}\label{sec:explicit}
The explicit electrostatic solver advances the plasma state according to the electrostatic Lorentz force and the Ampère-Maxwell equation assuming zero magnetic field
\begin{align}
    & \frac{\partial v_p}{\partial t}=\frac{qE_x}{m},\label{eq:lorentz} \\
    & \frac{\partial E_x}{\partial t}+4\pi J_x=0, \label{eq:amperes}
\end{align}
where $v_p,q,m$ is the (macro) particle velocity (in the $x$-direction), charge and mass respectively. Here $E_x$ is the electric field in the $x$-direction and $J_x$ represents the current density in the same direction.
The particle push, corresponding to \cref{eq:lorentz}, can be discretized using a forward finite difference scheme with time staggering
\begin{align*}
    v^{j+1/2}_{p} & = v^{j-1/2}_{p} + \frac{\Delta t qE^{j}_{x}(x^{j}_{p})}{m}, \\
    x^{j+1}_{p} & = x^{j}_{p} + \Delta t v^{j+1/2}_{p},
\end{align*}
where the upper indices denotes the timestep, while lower indices label the particles. The time staggering is realized through initializing $v_p$ half a timestep backwards in time as 
\begin{equation*}
    v^{-1/2}_{p} = v^{0}_{p} - \frac{\Delta tq E_x^{0}(x^{0}_{p})}{2m}.
\end{equation*}
The electric field is advanced as
\begin{align*}
    E_{x,i}^{j+1} = E_{x,i}^{j}-\Delta t 4\pi J_{x,i}^{j+1},
\end{align*}
where the lower indices denotes grid point. The current is interpolated to the grid from the particle position and momentum as 
\begin{align*}
    J_{x,i}^{j+1} = \sum_{p}^{N} w_pqS(x_{p}^{j+1}-x_i)v_p^{j+1/2}/V_g,
\end{align*}
where $N$ is the number of particles, $S$ is the shape function (1st order Cloud-In-Cell), $w_p$ is the macro-particle weight, $V_g$ is the cell volume and the lower index denotes particle index. The electric field is interpolated back to the particle position similarly
\begin{align*}
    E_{x}^{j}(x)=\sum^{N_g}_{i}S(x_i-x)E_{x,i}^{j},
\end{align*}
where $N_g$ are the number of adjacent grid nodes to $x$.

In short, this is how the methods in \texttt{pic\_solver} were used to execute the algorithm:

\begin{enumerate}
    \item \texttt{preStep():} Shift the particle velocities backward by $\Delta t/2$ to initialize the staggered leapfrog scheme.
    \item \texttt{preLoop():} Advance the electric field and reset the current density to zero.
    \item \texttt{processParticle():} Interpolate the electric field to the particle position, advance the particle, and deposit the current onto the grid.
    \item \texttt{postStep():} Advance the particle velocities by $\Delta t/2$ to remove the staggering.
\end{enumerate}

Note that, since neither the charge continuity equation nor Gauss’s equation is exactly satisfied in the numerical scheme, this discrepancy will eventually introduce errors in the simulation’s evolution. Furthermore, the plasma systems which can be realistically simulated using the electrostatic solver with periodic boundary conditions are restricted to systems with zero spatial average charge and current, as follows from integrating Gauss's law and \cref{eq:amperes} over a periodic region.

\begin{figure*}[b]
\centering
\includegraphics{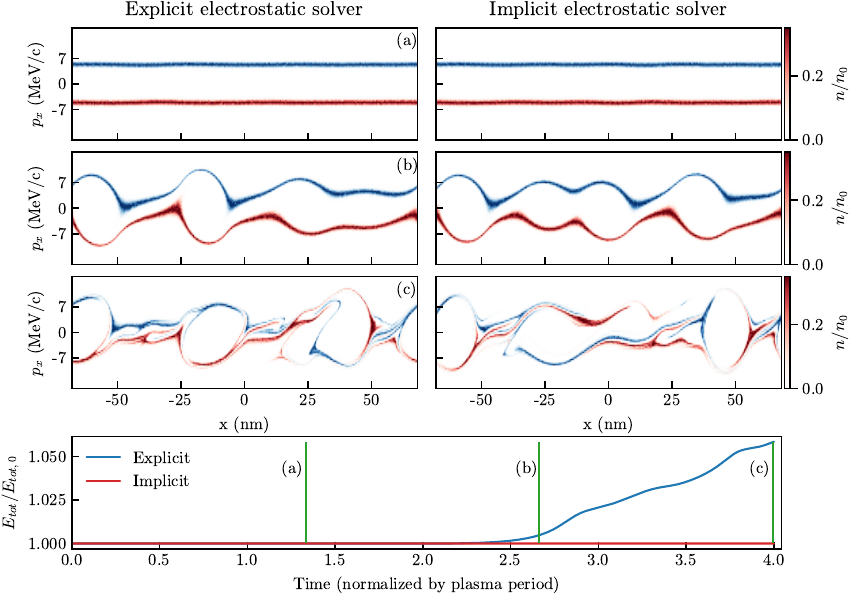}
\caption{Two-stream instability obtained by initializing two counter-propagating particle populations with equal density and opposite momentum. The benchmark was performed using an explicit electrostatic solver (left) and an implicit electrostatic solver (right). The explicit solver exhibits significant numerical heating at later times, whereas the implicit solver maintains a nearly constant total energy throughout the simulation.}
\label{fig:2stream}
\end{figure*}

\subsubsection{An implicit one-dimensional electrostatic solver}\label{sec:implicit}

The implicit electrostatic solver used in the following benchmark is based on the energy-conserving electrostatic PIC method presented by Markidis and Lapenta \cite{markidis_energy_2011}. In the original work, the nonlinear system is solved using a Jacobian-Free Newton--Krylov (JFNK) method. In the present implementation, we instead employ a fixed-point iteration in order to keep the solver compact and avoid introducing additional dependencies. The purpose of this implementation is not to provide a production-quality implicit solver, but rather to demonstrate how particle--field coupled updates can be implemented within the $\pi$-PIC framework. More advanced nonlinear solvers, including JFNK methods, can be incorporated through external C++ libraries.

The solver advances the same electrostatic system described in the previous section. However, instead of using a forward scheme, it discretizes the system \cref{eq:lorentz,eq:amperes} using midpoint integration:
\begin{align}
    & v_{p}^{j+1} = v_{p}^{j} + \Delta t\frac{q}{m}\bar{E}_{x}(\bar{x}_p), \nonumber \\ 
    & x_{p}^{j+1} = x^{j}_{p}+\Delta t\bar{v}_p, \nonumber \\
    & E_{x,i}^{j+1} = E_{x,i}^{j}- \Delta t 4\pi \bar{J}_{x,i}, \label{eq:imp}
\end{align}
where
\begin{align}
    & \bar{v}_p = (v_{p}^{j} + v_{p}^{j+1})/2, \nonumber \\
    & \bar{x}_p = (x^{j+1}_{p}+x^{j}_{p})/2. \nonumber \\
    & \bar{E}_{x}(\bar{x}_p) = \sum_{i}^{N_g} wS(\bar{x}_{p}-x_i)(E_{i}^{j+1}+E_{i}^{j})/2, \label{eq:imp_E} \\
    & \bar{J}_{x,i} = \sum_{p}^{N} qwS(\bar{x}_{p}-x_i)\bar{v}_p/V_g \label{eq:imp_J}.
\end{align}
The midpoint definitions of velocity and position can be used to eliminate $v_{p}^{n+1}$ and $x_{p}^{n+1}$, yielding the coupled system
\begin{align}
    & \bar{x}_p = x^{j}_{p} + \bar{v}_p\Delta t/2, \nonumber \\
    & \bar{v}_p = v_{p}^{j} + \frac{q}{m2}\bar{E}_{x}(\bar{x}_p)\Delta t, \nonumber 
\end{align}
which is evaluated in conjunction with \cref{eq:imp_J,eq:imp_E}, using an initial guess and iterating until convergence.

The entire nonlinear iteration is performed within the \texttt{postLoop()} method of the \texttt{pic\_solver}, where \texttt{ensemble::particleLoop()} is invoked to redistribute migrated particles according to their new cell assignment. Since particles and fields are updated concurrently during the implicit solve, the predefined particle-processing routine \texttt{processParticles()} cannot be used directly. At present, $\pi$-PIC does not provide a dedicated parallel execution structure for implicit solvers. 

\subsubsection{Benchmark using two-stream instability}

The two-stream instability is a commonly used benchmark for assessing numerical heating and energy conservation in particle-in-cell simulations. Here it is used to compare the conservation properties of the explicit and implicit electrostatic solvers implemented in $\pi$-PIC. \Cref{fig:2stream} shows the evolution of the instability at three different times together with the corresponding total energy of the system. Both solvers correctly reproduce the initial growth of the instability and the formation of phase-space structures. As the simulation progresses, however, clear differences emerge in their conservation properties.

After four plasma periods, the explicit solver exhibits a total energy increase of $5.85\%$, whereas the implicit solver shows a decrease of only $2.5\times10^{-3}\%$. This trend persists throughout the simulation. After twenty plasma periods, the total energy error of the implicit solver remains at the level of $4.3\times10^{-3}\%$, while the explicit solver has accumulated an energy increase of approximately $160\%$. 

The simulations shown used 200 particles per cell, 265 grid points, and $\Delta t = L_p/265$, where $L_p$ is the plasma period. The initial seeding of the instability depends on the spatial resolution and the number of particles per cell, causing the detailed evolution of the instability to vary between simulations with different resolutions. 

The superior energy conservation of the implicit solver is expected since the coupled system given by \cref{eq:imp,eq:imp_E,eq:imp_J} conserves energy when solved exactly \cite{markidis_energy_2011}. In contrast, the explicit discretization does not satisfy the corresponding discrete energy balance and therefore accumulates numerical heating over time. The benchmark demonstrates both the ability of $\pi$-PIC to accommodate implicit particle--field coupled algorithms and the usefulness of the framework for comparing different numerical formulations of the PIC method.

\subsection{Laser wakefield benchmark towards external codes}\label{sec:benchmark}

PIC codes are widely used for simulating laser wakefield acceleration (LWFA). Since its proposal \cite{tajima_laser_1979}, the acceleration technique has attracted much attention because it can sustain accelerating gradients which far exceed those produced by conventional accelerators \cite{esarey_physics_2009}, thereby enabling compact sources of relativistic particles. However, the complexity of the system makes analytical solutions largely intractable, rendering PIC simulations essential for investigating and understanding the underlying dynamics.

In this section, we will benchmark simulations of LWFA using the energy-conserving (EC) solver \cite{gonoskov_jcp_2024} and the Fourier-Boris (FB) solver of $\pi$-PIC, comparing it against the well-established PIC-code Smilei \cite{derouillat_smilei_2018}. We are not aiming at comparing the performance of these codes, because the approach used in EC is in an early development stage, lacking many significant improvements, such as higher order weighting and tailored field solvers. Instead, we focus on benchmarking and assessing capabilities of FB and EC solvers for LWFA, since in this system the benefits of energy conservation are less clear than that of numerical heating elimination in laser-solid interactions considered in \cite{gonoskov_jcp_2024}. In particular, we will focus on performance at low-resolution. 

The FB solver is a standard spectral solver with a Boris particle-pusher. For simulations with Smilei, we use the M4 Maxwell solver \cite{lu_time-step_2020} since it exhibits reduced numerical dispersion, which, when not corrected for, slows down the propagation of electromagnetic waves. The energy-conserving and FB solver in $\pi$-PIC uses a spectral solver, which inherently avoids this problem. 

\begin{figure}[h!]
    \centering
    \includegraphics[width=1.\columnwidth]{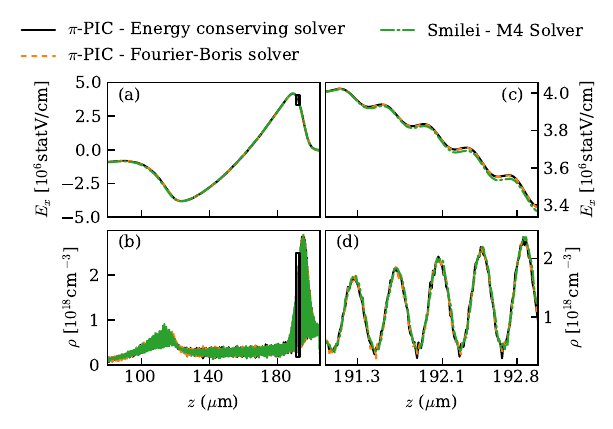}
    \caption{Longitudinal electric field (a, c) and plasma density (b, d) for 1D LWFA simulations using $\pi$-PIC with the EC and FB solver as well as Smilei with the M4 solver. The black rectangles in (a) and (b) indicate the enlarged area shown in (c) and (d) respectively.}
    \label{fig:1d}
\end{figure}

\begin{figure*}[t]
    \centering
    \includegraphics[width=0.75\textwidth]{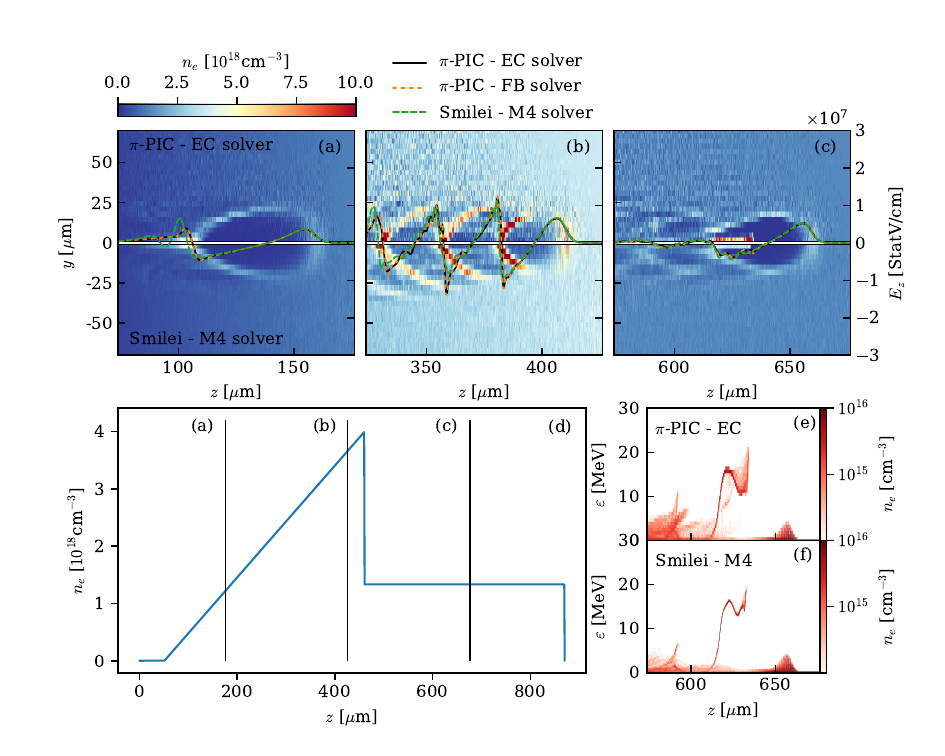}
    \caption{Panel (a-c) shows the on-axis density at three different positions in the plasma marked in panel (d). The upper half of panel (a-c) shows the $\pi$-PIC simulation and the lower half shows the Smilei simulation. Additionally the on-axis longitudinal electric field is shown for $\pi$-PIC using the EC solver (black) and FB solver (dashed orange) as well as the M4 solver of Smilei (green dashed). Panel (e-f) shows the phase-space at the longitudinal position corresponding to panel (c). }
    \label{fig:3d}
\end{figure*}

We begin by simulating the one-dimensional wakefield excitation that occurs when a laser interacts with a plasma density up-ramp. We employ a laser with wavelength $\lambda=\SI{1}{\micro\meter}$ and a Gaussian time-envelope spanning 30 laser cycles at full width half maximum (FWHM). The laser envelope has a peak amplitude $a_0=e|E|/(m_ec\omega)=4$, where $e$ denotes the elementary charge, $E$ the electric field amplitude, $m_e$ the electron mass, $c$ the speed of light and $\omega$ the laser frequency. The plasma density ramps from 0 to $4\cdot10^{18}\,\text{cm}^{-3}$ in $2^{10}\lambda=\SI{1024}{\micro\meter}$. The simulations have a space step $\Delta x=\lambda/16$, timestep $\Delta t=\lambda/(64c)$ and 8 particles per cell. 

The 1D simulations comparing the Smilei M4 solver and the $\pi$-PIC solver show excellent overall agreement (see \cref{fig:1d}). The only noticeable discrepancy is a minor difference in the buildup of the longitudinal electric field, most clearly visible in \cref{fig:1d}.(c). Note that this difference appears only in the comparison between the Smilei and $\pi$-PIC solvers, and not between EC and FB.

\begin{figure}[]
    \centering
    \includegraphics[width=0.72\columnwidth]{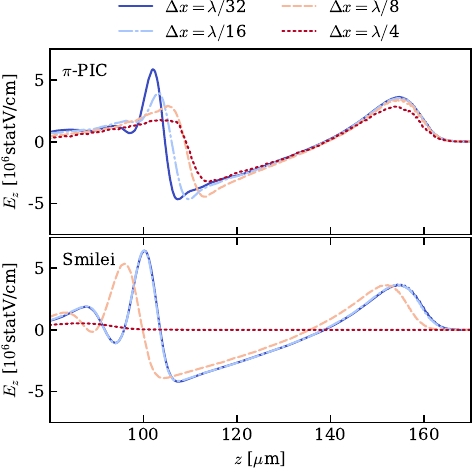}
    \caption{Shows the longitudinal electric field at the plasma upramp corresponding to position (a) in \cref{fig:3d} for simulations of $\pi$-PIC (upper) and Smilei (lower) for different resolutions.}
    \label{fig:res}
\end{figure}

Having shown satisfactory agreement in 1D, we move on to full 3D simulations of LWFA, where additionally injection and subsequent acceleration of particles is simulated. This is achieved via density downramp injection, using a plasma profile that decreases to one-third of the peak density ($n_e = 4 \cdot 10^{18}\,\text{cm}^{-3}$) at the end of the upramp, see \cref{fig:3d}.(d). The simulation parameters for the laser are identical to the 1D case, with the addition of laser focusing, characterized by a focus at the position of the downramp with a spot size of $40\lambda$. While the results in \cref{fig:3d} are qualitatively similar, the on-axis accelerating fields vary in amplitude and bubble-length between $\pi$-PIC solvers and Smilei, despite the wavefronts being identical. Moreover, there is a noticeable difference in the amount of injected charge and the resulting phase-space distribution of the accelerated beam.

The observed differences in particle injection suggest that the discrepancy between the solvers arises from their distinct macro-particle shape functions. The lowest order shape function available for the Smilei M4 solver in 3D is a triangular charge distribution corresponding to a 3-point stencil, whereas the EC and FB solvers use a Cloud-In-Cell shape function corresponding to a 2-point stencil. In addition, Smilei enforces exact charge conservation through Esirkepov's deposition scheme \cite{esirkepov_exact_2001}. In contrast, the current solvers in $\pi$-PIC do not implement a similar deposition scheme; however, the FB solver applies a Poisson-based charge correction. The agreement in plasma response between the EC and FB solvers indicates that effects of energy and charge conservation does not account for the differences observed between the $\pi$-PIC and Smilei solvers, pointing instead to the shape function as the most probable source of the discrepancy. 

To look further into these differences, we perform several simulations with different spatial and temporal resolutions, as shown in \cref{fig:res}. Since the FB and EC solvers produce similar results, we have excluded FB from the subsequent simulations. The timestep was $\Delta t=\Delta x/(4c)$ and $\Delta y = \Delta x = 4\lambda$. At high resolution ($\Delta x = \lambda/32$), the Smilei and EC solver show strong agreement at the front of the wake, but a qualitative difference emerges at the rear end of the wake. Additionally, Smilei converges to a solution faster with resolution, as the electric field for $\Delta x=\lambda/32$ and $\Delta x=\lambda/16$ is practically identical -- which is not the case for EC. 

For lower resolution there is a decrease in impact of the laser for both codes, shown at the front of the wake. This is to be expected, since the ponderomotive force which drives the wake, is an averaged force and by under-sampling the oscillations of the field the particles experience a lower effective push. For the EC solver there is an effective decrease of plasma wavelength for lower resolutions. Smilei, on the other hand, shows a decrease in propagation speed of the wake which becomes progressively worse with resolution. At a resolution of $n_x/\lambda=4$ (red line) the head of the wake is almost out of window. 

In conclusion, we find that the $\pi$-PIC EC and FB solvers retain a higher degree of accuracy at low resolution but the convergence of the code with higher resolution is weak compared to the Smilei M4 solver.

\section{Conclusion}
PIC methods provide powerful tools for studying plasmas across a wide range of environments and scales. The expansion of computing power makes previously unfeasible simulations achievable. Despite significant progress and ongoing research, challenges remain regarding numerical stability and conservation properties. To accommodate further research, we have developed a flexible PIC framework, which enables users to modify and extend the PIC scheme. 

The examples presented in this work demonstrate the flexibility of the framework through three distinct benchmarks: conservation-property tests using the two-stream instability, comparisons against established PIC codes in the context of laser wakefield acceleration, and absorbing-boundary validation using an extension. Additionally, we present extensions for handling of tight focusing and moving window. Together, these examples illustrate how new numerical methods and auxiliary functionality can be incorporated and evaluated within a common environment. More broadly, the examples presented here should be viewed as the first elements of a larger benchmark suite. Future benchmark categories may include particle-per-cell convergence studies, benchmarks of momentum, energy, and charge conservation, numerical-heating tests, boundary-condition validation, systematic comparisons of solvers using alternative field representations and particle--field coupling strategies, as well as validation of numerical methods for collisions and strong-field quantum electrodynamics. 

The current implementation is not without limitations. While the framework supports a broad range of particle--field coupling strategies and field representations, particle handling remains based on structured Cartesian meshes. Future developments may include a generalized particle infrastructure for unstructured mesh topologies, together with built-in support for alternative parallel particle-processing strategies and common data-access patterns employed by collisional solvers and methods with stronger particle–field coupling.

Ultimately, the purpose of $\pi$-PIC is not to promote a particular PIC algorithm, but to lower the barrier for developing and evaluating new ones. By separating simulation control, extensions, and solver implementation, the framework provides a common ground on which different numerical ideas can be tested, benchmarked, and compared under identical conditions.

\section*{Acknowledgment}
This work was supported by the Swedish Research Council Grant 2023-06998. Computational resources were provided by the National Academic Infrastructure for Supercomputing in Sweden (NAISS), partially funded by the Swedish Research Council through grant agreement no. 2022-06725.





\bibliographystyle{elsarticle-num}
\bibliography{refs.bib}







\end{document}